\documentclass[a4paper]{article}
\usepackage{a4,amsmath,cite}
\usepackage{dcolumn}
\usepackage{caption}

\begin{document}

\large
\title{\textbf{A new two-body relativistic potential model for pionic hydrogen}}

\author{D.A.
Kulikov\thanks{kulikov\_d\_a@yahoo.com},
R.S. Tutik\thanks{tutik@dsu.dp.ua}\\
\\
{\sl Theoretical Physics Department, Dniepropetrovsk National
University }\\
{\sl 72 Gagarin av., Dniepropetrovsk 49010, Ukraine} }

\date{}
\maketitle

\begin{abstract}
The new potential model for pionic hydrogen, constructed with the
employment of the two-body relativistic equation, is offered. The
relativistic equation, based on the extension of the $SL(2,C)$
group to the $Sp(4,C)$ one, describes the effect of the proton
spin and anomalous magnetic moment in accordance with the results
of the quantum electrodynamics. Within this approach, using the
experimental data on the strong energy level shift and width of
the $1s$ state in pionic hydrogen as input, the pion-nucleon
scattering lengths have been evaluated to be
$a_{\pi^{-}p}=0.0860(6)m_{\pi}^{-1}$ and
$a_{\pi^{0}n}=-0.1223(19)m_{\pi}^{-1}$.
\end{abstract}



\section{Introduction}

One of the most important sources of information on strong
interactions at low energy is the experiments performed with
hadronic atoms \cite{batty}. In particular, the data on the shifts
in energies and the widths in pionic \cite{schroder,PHC} and
kaonic \cite{beer} hydrogen are used to determine the threshold
parameters of the strong meson-nucleon S-matrix and the hadronic
scattering lengths. These investigations have been carried out
within the framework of the non-relativistic scattering theory
\cite{deser,trueman,ericson}, the effective field theory
techniques \cite{holstein,lyubovitskij,gasser,ivanov,borasoy} and
the potential models \cite{sigg,oades07,revai}.

The conventional potential model for pionic hydrogen, the bound
$\pi^{-}p$ system, is the one-particle approximation based on the
Klein-Gordon equation  \cite{sigg}. This model, however, does not
provide a consistent description of some effects. So, the
corrections raised by the proton spin and anomalous magnetic
moment are added to the electromagnetic energies of the $\pi^{-}p$
system ``by hand'', using their values obtained in the quantum
electrodynamics \cite{austen}. For the self-consistent description
of these effects in pionic hydrogen to be possible it is needed to
go beyond the one-particle approximation.

Up to now, the various approaches to deriving the relativistic
two-body equations describing the fermion-boson systems have been
offered \cite{tanaka,krolikowski,pilkuhn1,datta,kelkar}. Recently,
the extension of the $SL(2,C)$ group to the $Sp(4,C)$ one has been
proposed for this purpose, too \cite{kulikov,fb}. However, the
last treatment permits us to deal not only with the ordinary
Lorentz-scalar and Lorentz-vector potentials but also with the
Lorentz-tensor one, being responsible for the interaction with
anomalous magnetic moment.

The goal of the present work is to apply the above mentioned
approach, based on the extension of the $SL(2,C)$ group, for
constructing the model describing pionic hydrogen in the
self-consistent manner within the framework of the relativistic
two-body equation.

The outline of the Letter is as follows. In Section 2 we briefly
consider the relativistic two-body equation based on the extension
of the $SL(2,C)$ group and involving the Lorentz-scalar,
Lorentz-vector and Lorentz-tensor potentials. In Section 3 the
form of the potentials needed for the description of pionic
hydrogen is specified. In Section 4 correctness of the
corresponding two-body equation of the model is discussed. Section
5 deals with the electromagnetic binding energies for $\pi^{-}p$
system. Section 6 is devoted to the extraction of the hadronic
scattering lengths from the data on the strong energy level shift
and width of pionic hydrogen. Finally, conclusions are given in
Section 7.

Throughout the Letter we use the Minkowski metrics
$g^{mn}=\mathrm{diag}(1,-1,-1,-1)$ and the units in which
$\hbar=c=1$.

\section{Two-body equation with the extension
of the $SL(2,C)$ group} \label{part2}

It has been shown \cite{kulikov} that the extension of the
$SL(2,C) \equiv Sp(2,C)$ group to the $Sp(4,C)$ one permits us to
construct the wave equations for relativistic two-body systems.
Keeping in mind the description of pionic hydrogen, we consider
the system composed of a spin-1/2 fermion and a spin-0 boson. The
wave function of the system is represented by a Dirac spinor or,
in our treatment, by two $Sp(4,C)$ Weyl spinors $\varphi$ and
$\bar{\chi}$. The corresponding two-body wave equation without
interaction is a straightforward generalization of the
one-particle Dirac equation \cite{fb}
\begin{equation}\label{eq200}
P\bar{\chi}=(m_{+}+\tau^1\otimes I m_{-})\varphi, \qquad
\tilde{P}\varphi=(m_{+}+\tau^1\otimes I m_{-})\bar{\chi}\,
\end{equation}
where the $Sp(4,C)$ momentum spin-tensor, $P$, and its
conjugative, $\tilde{P}$, depend on the four-momenta of the
constituent particles of the system through the quantities
\begin{equation}\label{eq2}
w_m=\frac{1}{2}(p_{1m}+p_{2m}),\qquad
p_m=\frac{1}{2}(p_{1m}-p_{2m}).
\end{equation}
The mass parameters, $m_{\pm}$, are related to the masses of the
constituents by
\begin{equation}\label{eq3}
m_{+}=\frac{1}{2}(m_{1}+m_{2}),\qquad
m_{-}=\frac{1}{2}(m_{1}-m_{2}),
\end{equation}
whereas $I$ and $\tau^i$ stand for the unit $2\times 2$ matrix and
the Pauli matrices, respectively.

The above equation (\ref{eq200}) must be supplemented with the
subsidiary condition
\begin{equation}\label{eq4}
(w^m p_m -m_{+}m_{-}) \left(\begin{array}{c}
   \varphi \\
    \bar{\chi}
  \end{array}\right)\equiv \frac{1}{4}(p_1^2-p_2^2-m_1^2+m_2^2) \left(\begin{array}{c}
   \varphi \\
    \bar{\chi}
  \end{array}\right)=0,
\end{equation}
which guarantees that in the lack of the interaction the particles
are on the mass shell, being subjected to the free Dirac and
Klein-Gordon equations.



Because the wave equation (\ref{eq200}) with the subsidiary
condition (\ref{eq4}) describes two systems, which differ from
each other only in permutation of masses of the particles, we
accept, for definiteness, that the Dirac fermion (proton) has the
mass $m_1$ and the Klein-Gordon boson (pion) has the mass $m_2$.

As it has been shown \cite{fb}, the introduction of the
Lorentz-scalar, Lorentz-vector and Lorentz-tensor interaction
potentials into the free wave equation (\ref{eq200}) yields
\begin{eqnarray}\label{eq10}
&&[I\otimes \sigma^m (w_m+A_m)+\tau^1\otimes \sigma^m
(p_m+B_m)]\bar{\chi}
=[m_{+}+S_{+}+                                           \nonumber \\
&&\tau^1\otimes I (m_{-}+S_{-})
-\mathrm{i}I\otimes\sigma^m\tilde{\sigma}^n C_{mn}
-\mathrm{i}\tau^1\otimes\sigma^m\tilde{\sigma}^n D_{mn}]\varphi, \nonumber \\
 &&[I \otimes \tilde{\sigma}^m
(w_m+A_m)+\tau^1\otimes\tilde{\sigma}^m
(p_m+B_m)]\varphi=[m_{+}+S_{+}+ \rule{0pt}{7mm} \nonumber \\
&&\tau^1\otimes I (m_{-}+S_{-})
-\mathrm{i}I\otimes\tilde{\sigma}^m\sigma^n C_{mn}
-\mathrm{i}\tau^1\otimes\tilde{\sigma}^m\sigma^n
D_{mn}]\bar{\chi}.
\end{eqnarray}
Here the spin-tensors $P$ and $\tilde{P}$ are rewritten in terms
of the $2\times 2$ matrices $\sigma^m=(I,\, \boldsymbol\tau)$ and
$\tilde{\sigma}^m=(I,\, -\boldsymbol\tau)$ with
$\boldsymbol\tau=(\tau^1,\tau^2,\tau^3)$, the Lorentz-scalar
potentials, $S_{+}$, $S_{-}$, are involved through the mass
substitutions whereas the Lorentz-vector potentials, $A_m$, $B_m$,
and Lorentz-tensor ones, $C_{mn}$, $D_{mn}$, are involved through
the minimal and non-minimal substitutions on the four-momenta,
respectively.

Clearly, the obtained wave equation and the subsidiary condition
must be compatible, i.e., the operators in their left-hand sides
must commute. This can be achieved on assumption that the
potentials must depend on the relative coordinate
$x^m=x_{1}^m-x_{2}^m$ only through its transverse part
\begin{equation}\label{eq13}
x_{\bot}^m=(g^{mn}-w^m w^n/w^2)x_n
\end{equation}
with respect to the total four-momentum $w_m$, which is conserved
and so can be treated as the eigenvalue rather than the operator.

Furthermore, the subsidiary condition (\ref{eq4}) should be
subject to the above substitutions on the masses and four-momenta,
which should not change its form. All these result in the
following restrictions on the shape of the potentials
\begin{eqnarray}\label{eq11}
&&\omega^m\pi_m+\pi_m\omega^m=2w_m p^m, \qquad
M_{+}M_{-}+M_{-}M_{+}=2m_{+}m_{-},   \\
&&C_{mk}D^{mn}+D_{mk}C^{mn}=0, \qquad \omega^m
D_{mn}-D_{mn}\omega^m+ C_{mn}\pi^m - \pi^mC_{mn}=0 \nonumber
\end{eqnarray}
where $\omega_m=w_m+A_m$, $\pi_m=p_m+B_m$,
$M_{\pm}=m_{\pm}+S_{\pm}$.

Once these conditions have been satisfied, the wave equation
(\ref{eq10}) supplemented with the subsidiary condition
(\ref{eq4}) provides a quantum description of the two-body
problem, incorporating several important properties \cite{fb}. So,
it is manifestly covariant, has correct one-particle limits and
allows us to treat, in addition to the standard Lorentz-scalar and
Lorentz-vector potentials, also the interaction described by the
Lorentz-tensor potentials.

\section{Specific form of potentials for pionic hydrogen}
\label{part3}

Now let us specify the explicit form of the potentials needed for
the description of pionic hydrogen. The restrictions on this form
are as follows. Firstly, these potentials must obey the
compatibility conditions (\ref{eq11}). Secondly, the
electromagnetic interaction between the proton and the pion should
be described properly. This implies that the wave equation
(\ref{eq10}), being transformed into the semirelativistic
Schr\"odinger-like form, should produce such relativistic kinetic
energy, spin-orbit and Darwin terms which are consistent with the
``improved Coulomb potential'' of the quantum electrodynamics
\cite{austen}. Thirdly, when the proton is assumed to be much
heavier than the meson, Eq.~(\ref{eq10}) must be approximated by
the Klein-Gordon equation with the ordinary scalar and vector
potentials.

The Lorentz-vector potentials $A_m$ and $B_m$ satisfying the above
restrictions can be chosen in the form
\begin{equation}\label{eq16}
A_m=(\mathcal{F}-1)w_m, \qquad%
B_m=(\mathcal{F}^{-1}-1)p_m
-\frac{\mathrm{i}}{2\mathcal{F}^2}\frac{\partial
\mathcal{F}}{\partial x_{\bot}^m}\,,
\end{equation}
with $\mathcal{F}$ being a scalar function.

For the description of the electromagnetic interaction in pionic
hydrogen to be correct, we accept the parametrization of
$\mathcal{F}$ which follows the structure of the potential derived
by summing the contributions of the relevant diagrams in the
quantum electrodynamics \cite{sazdjian96}
\begin{equation}\label{eq16a}
\mathcal{F}=(1-2\mathcal{A}/E)^{1/2},
\end{equation}
where $E=2\sqrt{w^2}$ denotes the total energy and
$\mathcal{A}=\mathcal{A}(x_{\bot}^2)$ is the electrostatic
(Coulomb) potential.

For the Lorentz-scalar potentials $S_{+}$ and $S_{-}$, the
simplest solution to the compatibility conditions (\ref{eq11}) is
given by
\begin{equation}\label{eq17}
S_{\pm}=\frac{1}{2}\left((m_{+}+m_{-})^2+2m_{w}\mathcal{S}+\mathcal{S}^2\right)^{1/2}
\pm\frac{1}{2}\left((m_{+}-m_{-})^2+2m_{w}\mathcal{S}+\mathcal{S}^2\right)^{1/2}-m_{\pm},
\end{equation}
where
\begin{equation}\label{eq17a}
m_{w}=m_1 m_2/E\equiv (m_{+}^2-m_{-}^2)/E
\end{equation}
is the relativistic reduced mass, also referred to as the Todorov
variable \cite{todorov}, and $\mathcal{S}=\mathcal{S}(x_{\bot}^2)$
is the scalar function, which goes over to the scalar potential of
the Klein-Gordon equation in the limit when the proton is much
heavier than the meson.

It should be added that the Lorentz-vector potentials can be
chosen in either the electromagnetic-like form (\ref{eq16}) or the
time-like one. In the last case our Eq.~(\ref{eq10}) is
transformed into the equation by Kr\'{o}likowski
\cite{krolikowski} (for details see Ref.~\cite{fb}). However, for
the point Coulomb interaction such time-like potentials do not
lead to the correct relativistic recoil and Darwin terms in
contrast to the electromagnetic-like Lorentz-vector potentials
[see Eq.~(\ref{eq260}) below].

Notice that all different momentum-independent forms of the
Lorentz-scalar potentials, obeying the compatibility conditions,
can be brought to the form (\ref{eq17}) with the quantity
$\mathcal{S}$ thought to be energy dependent. But for our purposes
this energy dependence may be neglected because the binding energy
of mesic hydrogen is small compared to the proton and meson
masses.

In the following, only the manifestly covariant expressions
(\ref{eq16}) and (\ref{eq17}) for the Lorentz-vector and
Lorentz-scalar potentials will be used.

Concerning the Lorentz-tensor potentials $C_{mn}$ and $D_{mn}$, in
order to satisfy the compatibility conditions (\ref{eq11}), we
must put $C_{mn}=0$. Due to the matrix contents of
Eq.~(\ref{eq10}), the remaining potential $D_{mn}$ allows us to
describe the interaction with the anomalous magnetic moment in the
same form as the non-minimal coupling term introduced by Pauli
\cite{pauli} in the one-particle Dirac equation. By analogy with
the Pauli term, we set
\begin{equation}\label{eq18}
D_{mn}=\frac{k_1}{4m_1} \left( \frac{\partial A_n}{\partial
x_{\bot}^m}-\frac{\partial A_m}{\partial
x_{\bot}^n}+\frac{\partial B_n}{\partial
x_{\bot}^m}-\frac{\partial B_m}{\partial x_{\bot}^n} \right),
\end{equation}
where $k_1$ and $m_1$ denote the anomalous magnetic moment and the
mass of the fermion (for the proton, $k_1=1.793$).

\section{The two-body equation in the center-of-mass frame}
\label{part4}

Although the wave equation (\ref{eq10}) is relativistically
invariant, it is convenient, for practical purposes, to pass to
the center-of-mass frame. In this case we have
$\mathbf{w}\equiv(\mathbf{p}_1+\mathbf{p}_2)/2=0$, $E=2w_0$, so
that the subsidiary condition (\ref{eq4}) results in
$p_0=(m_1^2-m_2^2)/(2E)$. Then Eqs.~(\ref{eq4}) and (\ref{eq10})
are reduced to the Dirac-like form
\begin{equation}\label{eq24}
\begin{array}{l}
{\displaystyle
\left[\boldsymbol\tau\cdot\mathbf{p}+\frac{\mathrm{i}k_1 (E^2
\mathcal{F}^2-m_1^2+m_2^2)}{4m_1 E^2\mathcal{F}^2} \boldsymbol\tau
\cdot\boldsymbol\nabla\mathcal{A} \right]\phi} \\
{\displaystyle \phantom{aaaaaaaaaa}
=\left[E_1-\mathcal{A}+M_1\mathcal{F} +\frac{k_1}{2m_1
E\mathcal{F}^2}(\boldsymbol\tau\cdot[
\boldsymbol\nabla\mathcal{A}\times\mathbf{p}]) \right]\psi\,,}
\rule{0pt}{7mm} \\
{\displaystyle
\left[\boldsymbol\tau\cdot\mathbf{p}-\frac{\mathrm{i}k_1 (E^2
\mathcal{F}^2-m_1^2+m_2^2)}{4m_1 E^2\mathcal{F}^2} \boldsymbol\tau
\cdot\boldsymbol\nabla\mathcal{A} \right]\psi} \rule{0pt}{10mm} \\
 {\displaystyle \phantom{aaaaaaaaaa}
=\left[E_1-\mathcal{A}-M_1\mathcal{F} -\frac{k_1}{2m_1
E\mathcal{F}^2}(\boldsymbol\tau\cdot[
\boldsymbol\nabla\mathcal{A}\times\mathbf{p}]) \right]\phi\,,}
\rule{0pt}{7mm}
\end{array}
\end{equation}
where $\phi=\mathcal{F}^{-1/2}(1+\tau^1\otimes
I)(\bar{\chi}+\varphi)$ and
$\psi=\mathcal{F}^{-1/2}(1+\tau^1\otimes I)(\bar{\chi}-\varphi)$
are the components of the Dirac bispinor,
$M_1=(m_1^2+2m_w\mathcal{S}+\mathcal{S}^2)^{1/2}$,
$E_1=(E^2+m_1^2-m_2^2)/(2E)$, $m_1$ is the mass of the fermion and
$m_2$ is the mass of the boson.

Notice that, in the center-of-mass frame, the quantities
$\mathcal{S}$ and $\mathcal{A}$ which determine all the involved
potentials depend only on the distance between the particles,
$\sqrt{-x_{\bot}^2}=|\mathbf{x}|\equiv
|\mathbf{x}_1-\mathbf{x}_2|$. Therefore the spatial variables in
Eq.~(\ref{eq24}) can be separated in the same manner as for the
Dirac equation.

Thus, the derived equation (\ref{eq24}) is assumed as a basis of
our model for hadronic atoms. The correctness of its description
of relativistic effects can be demonstrated with following
limiting cases.

In the first case, it is supposed that the proton or, more likely,
spin-1/2 nucleus is much heavier than the pion, $m_{1}/m_2\gg 1$.
Then it is convenient to pass from the total energy $E$ to the
energy variable by Todorov \cite{todorov}:
\begin{equation}\label{eq25}
\epsilon_w=(E^2-m_1^2-m_2^2)/(2E)
\end{equation}
which together with the relativistic reduced mass $m_w$
corresponds to the relative motion satisfying, in the lack of
interaction, the effective one-body Einstein equation,
$\epsilon_w^2-m_w^2=\mathbf{p}^2$.

Expanding all terms of Eq.~(\ref{eq24}) in powers of $1/m_1$,
eliminating the ``small'' component, $\psi$, in favor of the
``large'' one, $\phi$, and using the identity
$(E_1-\mathcal{A})^2-m_1^{2}\mathcal{F}^2=(\epsilon_w-\mathcal{A})^2-m_w^2$,
Eq.~(\ref{eq24}) is reduced to the Klein-Gordon-like equation
\begin{equation}\label{eq26}
[\mathbf{p}^2+(m_w+\mathcal{S})^2-(\epsilon_w-\mathcal{A})^2]\phi%
=\frac{1}{m_1}[2\mathcal{A}(2m_w\mathcal{S}+\mathcal{S}^2)-\boldsymbol\nabla\mathcal{A}\cdot\mathbf{p}
+(\boldsymbol\tau\cdot[
\boldsymbol\nabla\mathcal{A}\times\mathbf{p}])]\phi
\end{equation}
that holds in the order of $1/m_1$.

The left-hand side of this equation is the usual Klein-Gordon
equation with the scalar and vector potentials, $\mathcal{S}$ and
$\mathcal{A}$, in which the Todorov variables $m_w$ and
$\epsilon_w$ appear instead of the reduced mass and energy,
whereas the right-hand side does indeed contain the three
corrections, raised by the heavy spin-1/2 particle, which describe
the interference between the scalar and vector potentials, the
Darwin term and the spin-orbit interaction, respectively.

The second limiting case is the semirelativistic approximation
which is better suited for discussing the lowest-order
relativistic corrections for pionic hydrogen. Now it is supposed
that the meson mass is of the same order as the proton mass, but
the binding energy, $\epsilon=E-m_1-m_2$, is much smaller than
these masses, $\epsilon/m_{1,2}\ll 1$. Since the magnitudes of the
potentials must also be small as compared to $m_{1,2}$, only the
electrostatic potential, and not the strong interaction one, can
be treated in this way.

Then, in the lack of the Lorentz-scalar (strong) interaction,
after performing the expansion in powers of $1/m_{1,2}$,
Eq.~(\ref{eq24}) is transformed into
\begin{eqnarray}\label{eq260}
\left(\frac{\mathbf{p}^2}{2\mu}-\epsilon+\mathcal{A}\right)\phi=
\left\{ \frac{1}{2\mu}\left[
\left(1-\frac{3\mu}{M}\right)\epsilon^2
-2\left(1-\frac{\mu}{M}\right)\epsilon\mathcal{A}+\mathcal{A}^2
\right] \right. \phantom{Gggggggg}  \\
\left.-\frac{1}{4m_1^2}\left(k_1+\frac{1}{2}+\frac{m_1}{m_2}\right)\boldsymbol\nabla^2\mathcal{A}
-\frac{1}{2m_1\mu}\left(k_1+\frac{1}{2}+\frac{m_1}{2M}\right)(\boldsymbol\tau\cdot[
\boldsymbol\nabla\mathcal{A}\times\mathbf{p}]) \right\}\phi\,,
\nonumber
\end{eqnarray}
where $M=m_1+m_2$, $\mu=m_1 m_2/M$.

Putting $\mathcal{A}$ as the point Coulomb potential, we obtain
just the same relativistic kinetic energy, spin-orbit and Darwin
terms as those derived with the `improved Coulomb potential' in
the quantum electrodynamics \cite{austen}.

Thus, we may conclude that the proposed relativistic two-body
equation (\ref{eq24}) with the electromagnetic interaction
introduced through the Lorentz-vector and Lorentz-tensor
potentials, (\ref{eq16}) and (\ref{eq18}), correctly describes the
relativistic corrections for pionic hydrogen.

\section{Electromagnetic energies}
\label{part5}

For evaluating the stationary electromagnetic binding energies of
pionic hydrogen, we use the derived equation (\ref{eq24}) without
the strong interaction. As it has been shown, the electromagnetic
interaction is introduced with the Lorentz-vector and
Lorentz-tensor potentials. According to Eqs.~(\ref{eq16}) and
(\ref{eq18}), these potentials are determined by the single
potential $\mathcal{A}$, which we represent as the sum of three
parts
\begin{equation}\label{eq300}
\mathcal{A}_{\mathrm{e.m.}}=\mathcal{A}_{\mathrm{pc}}+\mathcal{A}_{\mathrm{ext}}+\mathcal{A}_{\mathrm{vac}}.
\end{equation}

Here $\mathcal{A}_{\mathrm{pc}}=-\alpha/r$ is the point Coulomb
potential, $\alpha=1/137.036$;
\begin{equation}\label{eq30}
\mathcal{A}_{\mathrm{ext}}=-\frac{\alpha}{r}\,(\mathrm{erf}\,(r/s)-1),
\qquad s^2=\frac{2}{3}(\langle r_{p}^2\rangle+\langle
r_{\pi}^2\rangle)
\end{equation}
is the addition caused by the finite extension of the Gaussian
charge distributions with $\langle r_{p}^2\rangle$ and $\langle
r_{\pi}^2\rangle$ being the r.m.s. charge radii of the proton and
pion;
\begin{equation}\label{eq301}
\mathcal{A}_{\mathrm{vac}}=\frac{\alpha}{\pi}\int_0^1 \mathrm{d}v
\frac{v^2 (1-v^2
/3)}{1-v^2}(\mathcal{A}_{\mathrm{pc}}(r)+\mathcal{A}_{\mathrm{ext}}(r))e^{-2m_e
r/\sqrt{1-v^2}}
\end{equation}
is the Uehling potential, commonly used for describing the vacuum
polarization at $O(\alpha^2)$ in hadronic atoms, which is smeared
at short distances due to the finite extension of the charge
distributions ($m_e$ is the electron mass).

Eqs.~(\ref{eq24}) with these potentials were solved numerically
with the shooting method. For the pion charge radius, the value
$\sqrt{\langle r_{\pi}^2 \rangle }= 0.663(6)$ fm deduced from the
$\pi e$ scattering experiments \cite{amendolia} was used. The
proton charge radius, $\sqrt{\langle r_{p}^2 \rangle}= 0.8750(68)$
fm, and the masses of particles were taken from the PDG tables
\cite{pdg2008}. Using these values, the electromagnetic binding
energy of the $1s$ state in pionic hydrogen is found to be
$\epsilon_{1s}=-3238.252$ eV with an uncertainty of $0.002$ eV due
to the uncertainties in the proton and pion charge radii.

In order to compare our result with those reported in literature,
it is convenient to consider the difference between the
electromagnetic binding energies obtained within the framework of
our two-body description and the Klein-Gordon equation, which
proved to be $\Delta\epsilon^{\mathrm{e.m.}}_{1s}=-3.092$ eV. In
fact, $\Delta\epsilon^{\mathrm{e.m.}}_{1s}$ is the sum of the
corrections due to the three different effects, namely, the vacuum
polarization effect, the finite size effect and the effect of the
relativistic recoil, the proton spin and the anomalous magnetic
moment.

\begin{table}
\caption{ Contributions  to the correction
$\Delta\epsilon^{\mathrm{e.m.}}_{1s}$ to the pionic hydrogen
binding energy in eV. Numbers in brackets are the uncertainties
due to uncertainties in the proton and pion charge radii.}
\vspace{0.5cm} \label{tab:1}
\begin{center}
\begin{tabular}{|p{5.2cm}|p{1.8cm}|p{1.8cm}|p{1.8cm}|p{1.8cm}|}
\hline
 & This work & Ref.~\cite{sigg} &  Ref.~\cite{lyubovitskij} &  Ref.~\cite{kelkar} \\
\hline
Finite size effect & \hspace*{0.02em} 0.106(3) & \hspace*{0.02em} 0.102(3)  & \hspace*{0.02em} 0.100 & \hspace*{0.02em} 0.102(9) \\
Vacuum polarization, order $O(\alpha^2)$ & -3.245 & -3.246  & -3.241 &  \\
Relativistic recoil, proton spin and anomalous magnetic moment &
$\phantom{oooooooooooo}$ \hspace*{0.02em} 0.047 &
$\phantom{oooooooooooo}$ \hspace*{0.02em} 0.047  \cite{austen} &
$\phantom{oooooooooooo}$ \hspace*{0.02em} 0.047 & $\phantom{oooooooooooo}$ \hspace*{0.02em} 0.0417 \\
Vacuum polarization, order $O(\alpha^3)$ & & -0.018  &  &  \\
Vertex correction &  & \hspace*{0.02em} 0.007  &  &  \\
\hline
\end{tabular}
\end{center}
\end{table}

In Table \ref{tab:1}, we compare the corrections to the
electromagnetic binding energy of pionic hydrogen obtained with
the present approach and with the other methods
\cite{sigg,lyubovitskij,kelkar}.

It should be stressed that in Ref.~\cite{sigg} the finite size
correction was evaluated within the Klein-Gordon equation by
employing the Gaussian charge distribution, whereas the correction
due to the relativistic recoil, the proton spin and the anomalous
magnetic moment evaluated for the point Coulomb potential
\cite{austen} has been added ``by hand''. In
Ref.~\cite{lyubovitskij}, the electromagnetic corrections were
calculated within the framework of the chiral perturbation theory
using the same input data as in Ref.~\cite{sigg}. Notice that
nearly all the difference between our result and the results of
these two approaches comes from the values of the r.m.s. proton
and pion charge radii. In Ref.~\cite{kelkar}, the finite size
effect and the relativistic correction were taken into account by
employing the Breit-type equation. In contrast to
Refs.~\cite{austen,lyubovitskij} as well as to our conclusion, the
contribution of the proton anomalous magnetic moment in
Ref.~\cite{kelkar} is found to be negligible that slightly reduces
the total relativistic correction, as seen from Table \ref{tab:1}.
This suppression seems to be caused by the sharp decrease of the
hadron form factors, used in Ref.~\cite{kelkar}, at small $r$ (or
at large momentum transfer), whereas the anomalous magnetic moment
reveals itself only in this region.

Thus, we may conclude that the calculated corrections to the
electromagnetic binding energy of pionic hydrogen are in good
agreement with those quoted in literature. However, it should be
stressed that, in contrast to the other approaches, the proposed
relativistic two-body equation allows us to evaluate the finite
size effect and the effect of the relativistic recoil, the proton
spin and anomalous magnetic moment in the self-consistent way
without resorting to the semirelativistic expansions in powers of
$1 /c^2$. We anticipate that for heavier mesic atoms the developed
approach may give the more distinguished values of the finite size
corrections as compared to the other approaches, because the
charge radii of the nuclei are larger.

\section{Pion-nucleon scattering lengths}
\label{part6}

Now we are going to incorporate the strong interaction potential
into the consideration and to calculate the pion-nucleon
scattering lengths, using the existing experimental data on the
strong energy-level shift and width of pionic hydrogen.

The extraction of the pion-nucleon scattering lengths will be done
in the isospin-symmetrical single-channel approximation when the
electromagnetic interaction is completely switched off and, by
convention, the masses of the pions and nucleons are taken equal
to the physical masses of the charged pion and proton,
respectively. The detailed discussion of the influence of the
isospin-breaking effects on the meson-nucleon scattering
parameters can be found in Refs.~\cite{gasser,oades07}.

Our relativistic model of pionic hydrogen utilizes the following
potentials
\begin{equation}\label{eq50s}
\mathcal{A}=\mathcal{A}_{\mathrm{e.m.}}, \qquad
\mathcal{S}=U_{\mathrm{str}}\,,
\end{equation}
where it is assumed that the (optical) strong interaction
potential $U_{\mathrm{str}}$ is the Lorentz-scalar and has the
square-well form
\begin{equation}\label{eq43}
U_{\mathrm{str}}(r) =\left\{
\begin{array}{cl}
U_{0}  &  \mathrm{for} \;\, r\leq \sqrt{\frac{5}{3}}\, r_0\,, \\[0.2cm]
0 & \mathrm{otherwise}\,.
\end{array}
\right.
\end{equation}

For fitting the position and width of the quasistationary energy
level of the $1s$ state in pionic hydrogen, Eqs.~(\ref{eq24}) were
solved numerically with varying the complex value of the strong
interaction potential strength, $U_0$. Under realization of this
procedure, the most recent experimental data by the PSI
collaboration, $\epsilon^{\mathrm{str}}_{1s} = -7.120 \pm 0.008$
(stat) $\pm 0.006$ (syst) eV and $\Gamma_{1s} = 0.823 \pm 0.018$
eV \cite{PHC}, which are consistent with, but more precise than,
the earlier ones \cite{schroder}, have been used. For adapting the
experimental data to our single-channel approximation, the strong
decay channel ($\pi^{-}p\rightarrow\pi^{0}n$) was separated out
with the replacement of the total decay width, $\Gamma_{1s}$, by
the partial width, $\Gamma_{1s}^{\pi^{0}n}=\Gamma_{1s}/(1+P^{-1})$
where $P=1.546(9)$ is the Panofsky ratio \cite{spuller}.

After the potential strength, $U_0$, had been adjusted, the
electromagnetic interaction was switched off and the radial
equations were solved once again to produce the pion-nucleon
scattering length \cite{sigg}. The whole procedure was repeated
with varying the parameter $r_0$ between $0.5$ fm and $1.5$ fm.

The calculations with the exponential and Gaussian potentials
instead of the square-well potential did not changed the values of
$a^{\mathrm{h}}$, as it is expected for the effective-range
theory.

The final results for the pion-nucleon scattering lengths in our
models is
\begin{equation}\label{eq52}
 a_{\pi^{-}p}=0.0860(6)m_{\pi}^{-1}\,, \qquad a_{\pi^{0}n}=-0.1223(19)m_{\pi}^{-1}\,.
\end{equation}
It is to be pointed that the error in $a_{\pi^{-}p}$  arises
mainly from the vague value of $r_0$ ($0.5$ fm $\leq r_0 \leq$
$1.5$ fm) and is reduced to $0.0002m_{\pi}^{-1}$ if the value
$r_0=1.0$ fm is fixed, whereas almost all the error in
$a_{\pi^{0}n}$ comes from the experimental uncertainty in
$\Gamma_{1s}$.

\section{Conclusion}
\label{part7}

The main object of the present work was to construct the new
potential model for describing pionic hydrogen within the
framework of the relativistic two-body equation incorporating in
the self-consistent manner the effects of the proton spin and
anomalous magnetic moment.

With using the proposed model, the electromagnetic binding energy
of pionic hydrogen has been evaluated without resorting to the
semirelativistic expansions in powers of $1 /c^2$. This enabled us
to treat non-perturbatively the finite extension of the charge
distributions of the proton and meson. We have found that in the
presence of the Gaussian charge distributions the corrections to
the binding energy due to the relativistic recoil, the proton spin
and the anomalous magnetic moment almost coincides with those
obtained for the point Coulomb potential \cite{sigg,austen}. This
is in contrast to Ref.~\cite{kelkar}, in which the contribution of
the anomalous magnetic moment has proved to be negligible.

Furthermore, the pion-nucleon scattering lengths have been
extracted from the experimental data on the strong energy level
shift and width of the $1s$ state in pionic hydrogen. It turns out
that the effect of the proton spin and anomalous magnetic moment
on the hadronic scattering lengths is exceedingly small. For
instance, switching off the proton anomalous magnetic moment
subtracts approximately $0.5\times 10^{-4} m_{\pi}^{-1}$ from
$a_{\pi^{-}p}$ that exceeds the precision of the experimental
data. Notice that our results on the pion-nucleon scattering
lengths are consistent with recent extractions within the
framework of other potential models. In particular, values in
Eqs.~(\ref{eq52}) agree reasonably well with quantities
$a_{\pi^{-}p}=0.0859(6)m_{\pi}^{-1}$, $a_{\pi^{0}n}
=-0.1243(15)m_{\pi}^{-1}$, calculated using the three-channel
relativized potential model \cite{oades07}, and
$a_{\pi^{-}p}=0.0870(5)m_{\pi}^{-1}$, $a_{\pi^{0}n}
=-0.125(4)m_{\pi}^{-1}$, calculated using the non-relativistic
scattering theory \cite{ericson}. The agreement would be even
better if we used, as in Refs.~\cite{oades07,ericson}, the earlier
value for the pionic hydrogen width \cite{schroder} which is
substantially larger than the latest one \cite{PHC}.

It is expected that the relativistic effects will be more profound
for kaonic hydrogen because its properties are strongly influenced
by the existence of the $\Lambda(1405)$ resonance.


\section*{Acknowledgments}

We thank Dr. J. R\'evai for discussing the details of his work
\cite{revai}. This research was supported by a grant N 0106U000782
from the Ministry of Education and Science of Ukraine which is
gratefully acknowledged.



\begin{thebibliography}{00}


\bibitem{batty} C.J. Batty, E. Friedman, A.
Gal, {\rm Phys. Rept.} \textbf{287}, 385 (1997).






\bibitem{schroder} H.-Ch. Schr\"oder {\it  et al.}, {\rm Eur. Phys. J. C } \textbf{21} (2001) 473.


\bibitem{PHC} L.M. Simons [Pionic Hydrogen Collaboration], {\rm Proceedings of International Workshop on Exotic Hadronic Atoms,
Deeply Bound Kaonic Nuclear States and Antihydrogen: Present
Results, Future Challenges, Trento, 2006}, edited by C. Curceanu,
A. Rusetsky and E. Widmann, p.8 [arXiv:hep-ph/0610201].




\bibitem{beer} G. Beer {\it  et al.}, {\rm Phys. Rev. Lett.} \textbf{94} (2005) 212302.



\bibitem{deser} S. Deser, M.L. Goldberger, K. Baumann, W. Thirring, {\rm Phys.
Rev.} \textbf{96} (1954) 774.

\bibitem{trueman} T.L. Trueman, {\rm Nucl. Phys.} \textbf{26} (1961) 57.

\bibitem{ericson} T.E.O. Ericson, B. Loiseau, S. Wycech, {\rm Phys. Lett. B}
\textbf{594} (2004) 76.



\bibitem{holstein} B.R. Holstein, {\rm Phys.
Rev. D} \textbf{60} (1999) 114030.

\bibitem{lyubovitskij} V.E. Lyubovitskij, A. Rusetsky, {\rm Phys. Lett. B}
\textbf{494} (2000) 9. 

\bibitem{gasser} J. Gasser, M.A. Ivanov, E. Lipartia, M. Moj\v{z}i\v{s}, A. Rusetsky, {\rm Eur. Phys.
J. C} \textbf{26} (2002) 13. 

\bibitem{ivanov} A.N. Ivanov, M. Faber, A. Hirtl, J. Marton, N.I. Troitskaya, {\rm Eur. Phys.
J. A} \textbf{18} (2003) 653. %

\bibitem{borasoy}
B. Borasoy, R. Ni{\ss}ler, W. Weise,  {\rm Phys. Rev. Lett.}
\textbf{94} (2005) 213401.




\bibitem{sigg} D. Sigg, A. Badertscher, P.F.A. Goudsmit, H.J. Leisi,
G.C. Oades, {\rm Nucl. Phys. A} \textbf{609} (1996) 310.

\bibitem{oades07} G.C. Oades, G. Rasche, W.S. Woolcock,
E. Matsinos, A. Gashi,  {\rm Nucl. Phys. A} \textbf{794} (2007)
73.

\bibitem{revai} J. R\'evai, N.V. Schevchenko, {\rm Few Body Syst.}
\textbf{42} (2008) 83.




\bibitem{austen} G.J.M. Austen, J.J. de Swart,
{\rm Phys. Rev. Lett.}  \textbf{50} (1983) 2039.



\bibitem{krolikowski} W. Kr\'{o}likowski, {\rm Acta Phys. Pol. B}
\textbf{10} (1979) 739.



\bibitem{pilkuhn1} H. Pilkuhn, {\rm J. Phys. B}  \textbf{25} (1992) 299.

\bibitem{tanaka} T. Tanaka, A. Suzuki, M. Kimura, {\rm Z. Phys. A}
\textbf{353} (1995) 79.

\bibitem{datta} S.N. Datta, A. Misra,
{\rm J. Chem. Phys. } \textbf{125} (2006) 084111.

\bibitem{kelkar} N.G. Kelkar, M. Nowakowski,
{\rm Phys. Lett. B} \textbf{651} (2007) 363.




\bibitem{kulikov} D.A. Kulikov, R.S. Tutik, A. P. Yaroshenko,
{\rm Phys. Lett. B} \textbf{644} (2007) 311.

\bibitem{fb} D.A. Kulikov, R.S. Tutik, {\rm Mod. Phys. Lett. A}  \textbf{23} (2008) 1829.



\bibitem{todorov} I.T. Todorov, {\rm Phys. Rev. D}
\textbf{3} (1971) 2351.

\bibitem{pauli} W. Pauli, {\rm Rev. Mod. Phys.}  \textbf{13} (1941) 203.

\bibitem{sazdjian96} H. Jallouli, H. Sazdjian, {\rm Phys. Lett. B}
\textbf{366} (1996) 409.










\bibitem{amendolia}
S. Amendolia {\it et al.},  {\rm Nucl. Phys. B}  \textbf{277}
(1986) 168.

\bibitem{pdg2008}
C. Amsler {\it et al.},  {\rm Phys. Lett. B} \textbf{667} (2008)
1.

\bibitem{spuller}
J. Spuller {\it et al.},  {\rm Phys. Lett. B} \textbf{67} (1977)
479.

\end{thebibliography}
\end{document}